\documentstyle[12pt]{article}
\newcommand{\be}{\begin{equation}}
\newcommand{\ee}{\end{equation}}
\newcommand{\bea}{\begin{eqnarray}}
\newcommand{\eea}{\end{eqnarray}}
\newcommand{\ba}{\begin{array}}
\newcommand{\ea}{\end{array}}

\def\bbox{{\,\lower0.9pt\vbox{\hrule \hbox{\vrule height 0.2 cm
\hskip 0.2 cm \vrule height 0.2 cm}\hrule}\,}}
\newcommand{\dsl}{\pa \kern-0.5em /}

\font\mybb=msbm10 at 12pt
\def\bb#1{\hbox{\mybb#1}}

\def\bH {\bb{H}}

\def\appendix#1{
  \addtocounter{section}{1}
  \renewcommand{\thesection}{\Alph{section}}
  \section*{Appendix \thesection\protect\indent \parbox[t]{11.15cm}
  {#1} }
  \addcontentsline{toc}{section}{Appendix \thesection\ \ \ #1}
  }

\begin{document}

%%%%%%%%%%%%%%%% title page %%%%%%%%%%%%%%%%%%%%%%%%%%%%%%%%%%%%

%%%%%%%%%%%%%%%% title page %%%%%%%%%%%%%%%%%%%%%%%%%%%%%%%%%%%%

\begin{titlepage}
\rightline{DAMTP-2010-54}
\rightline{UG-2010-74}
\rightline{MIT-CTP-4166}
%\begin{flushright}
%\end{flushright}

\vfill

\begin{center}
\baselineskip=16pt {\Large\bf  Gravitons in Flatland\footnote{Talk given by PKT at the workshop {\it Cosmology, the Quantum Vacuum and Zeta Functions} in celebration of the 60th birthday of Emilio Elizalde; Barcelona, 8-10 March, 2010.} }
%{$^\star$}}
\vskip 0.3cm
{\large {\sl }}
\vskip 10.mm
{\bf Eric Bergshoeff$^\dagger$,  Olaf Hohm$^*$  and  Paul K.
Townsend$^+$}
\vskip 1cm
{\small
$^\dagger$
Centre for Theoretical Physics, University of Groningen, \\
Nijenborgh 4, 9747 AG Groningen, The Netherlands \\
}
\vspace{6pt}
{\small
  $^*$
Center for Theoretical Physics, 
Massachusetts Institute of Technology, \\
Cambridge, MA 02139, USA \\
}
\vspace{6pt}
{\small
  $^+$
Department of Applied Mathematics and Theoretical Physics\\
Centre for Mathematical Sciences, University of Cambridge, \\
Wilberforce Road, Cambridge, CB3 0WA, UK\\
}
\end{center}
\vfill

\par
\begin{center}
{\bf ABSTRACT}
\end{center}
\begin{quote}

We review some features of three-dimensional (3D) massive gravity theories. In particular, we stress the role of the Schouten tensor, explore an analogy with Lovelock gravity and discuss renormalizabilty.

\vfill

\end{quote}
\end{titlepage}
%%%%%%%%%%%%%%%%%%%%%%%%%%%%%%%%%%%%%%

%\section*{}
%\label{sec:1}

Edwin Abbot's 1884 novella  {\it Flatland}  \cite{Flatland} was the first of many explorations of a hypothetical world in which there are only two space dimensions (although Charles Hinton's  {\it A Plane World}, also published in 1884, may deserve equal credit\footnote{We thank Gary Gibbons for bringing Hinton's work  to our attention.}). In the context of special relativity, Flatland may be considered to be a three-dimensional (3D) Minkowski spacetime. Of course, the term ``Flatland'' might  be considered inappropriate in the context of General Relativity, but the 3D Einstein equations imply that the spacetime curvature is entirely determined by the matter content,  so spacetime is still locally flat  outside sources \cite{Staruszkiewicz:1963zz,Deser:1983tn,Giddings:1983es}; this means that 3D GR does not  admit  gravitational waves, and there are consequently no massless gravitons. However, it is possible to add higher-order terms (in a derivative expansion) to the Einstein-Hilbert (EH) action such that when the action is expanded  about the Minkowski vacuum (in powers of the metric perturbation) the quadratic approximation to it is a Minkowski space field theory propagating  {\it massive} gravitons.  This contribution reviews the current status of these 3D ``massive gravity''  theories, focusing  on the small fluctuations about Minkowski space: i.e. on gravitons in Flatland. 

We start from the 3D Fierz-Pauli (FP) equations for a totally symmetric rank-$s$ tensor $\varphi^{(s)}$, which is also traceless for $s\ge2$. These equations consist of  the dynamical equation 
\begin{equation}
 \left(\Box -m^2 \right)\varphi^{(s)}  =0
 \end{equation}
together with the  subsidiary condition
 \begin{equation}\label{subsid}
 \partial^\nu \varphi^{(s)}_{\nu\mu_2\cdots \mu_s} =0\, . 
\end{equation}
They  are equivalent to the single equation 
\begin{equation} \label{FP}
\left[ \bH  - sm \right] \left[ \bH + sm \right]\varphi^{(s)} =0\, , 
\end{equation}
where $m$ is a mass parameter and  $\bH$ is the spin-$s$ unit-mass ``helicity operator'' :
\begin{equation}
\bH_{\mu_1\dots \mu_s}{}^{\nu_1\dots\nu_s} = 
s \varepsilon_{(\mu_1}{}^{\tau (\nu_1} \delta_{\mu_2}^{\nu_2}\cdots \delta_{\mu_s)}^{\nu_s)} \, \partial_\tau\, . 
\end{equation}

It is evident from (\ref{FP}) that two spin-$s$ modes are propagated, of helicities $\pm s$. The generalized equation 
\begin{equation}\label{genFP}
\left[ \bH  - sm_+ \right] \left[ \bH + sm_- \right]\varphi^{(s)} =0
\end{equation}
also propagates two modes of helicities $s$ and $-s$ but with, respectively,  masses $m_+$ and $m_-$ (both of which may assumed to be positive). Since parity flips the sign of helicity, it follows that parity is violated whenever $m_+ \ne m_-$. In particular, we may take $m_-\to \infty$ for fixed 
$m_+ = \mu$, in which case the second-order equation (\ref{genFP}) degenerates to the first-order ``self-dual'' 
equation \cite{Townsend:1983xs,Aragone:1986hm,Tyutin:1997yn}
\begin{equation}\label{self-dual}
\left[ \bH  - s\mu \right] \varphi^{(s)} =0\, . 
\end{equation}
Note that this still implies the subsidiary condition (\ref{subsid}). 

{}For any of the above massive spin-$s$  field equations, there is a systematic procedure for finding an equivalent set of  equations for a spin-$s$ {\it gauge theory} \cite{Bergshoeff:2009tb}. We first  relax the tracelessness condition on $\varphi$ (in the case that $s\ge2$) and then solve the subsidiary condition by writing
\begin{equation}\label{einstein-s}
\varphi^{(s)}_{\mu_1\cdots\mu_s} =  - \frac{1}{s!} \varepsilon_{\mu_1}{}^{\tau_1\nu_1}\cdots 
\varepsilon_{\mu_s}{}^{\tau_s\nu_s}\partial_{\tau_1}\cdots\partial_{\tau_s} h_{\nu_1\cdots \nu_s} 
\equiv G_{\mu_1\cdots\mu_s}  
\end{equation}
for a rank-$s$ gauge potential $h$, with rank-$s$  field-strength $G$ invariant under  the gauge transformation\footnote{For spin $s\ge3$ fields in 4D it is usual to assume a weaker gauge invariance in which the parameter is traceless, in which case  the  field strength is always second-order in derivatives.}
\begin{equation}
h_{\mu_1\cdots \mu_s}  \to h_{\mu_1\cdots \mu_s}  + \partial_{(\mu_1} \xi_{\mu_2\cdots\mu_s)}
\end{equation}
for arbitrary rank-$(s-1)$ symmetric tensor parameter $\xi$. The subsidiary condition is then replaced by the Bianchi-type identity
\begin{equation}
\partial^\nu G_{\nu\mu_1\cdots \mu_{s-1}} \equiv 0\, , 
\end{equation}
but the tracelessness condition on $\varphi$, which we initially relaxed, must now be re-imposed as a condition on $G$.

Applied to the ``self-dual'' equation (\ref{self-dual}) for $s=1$,  this procedure yields the field equations of ``topologically massive electrodynamics'', as has been known for some time \cite{Jackiw:1990ka}.  The application to the same self-dual equation for $s=2$ was presented in \cite{Andringa:2009yc}; in this case we have the equations
\begin{equation}\label{spin2}
\left[ \bH  - 2\mu \right] G=0\, , \qquad \eta^{\mu\nu}G_{\mu\nu} =0\, , 
\end{equation}
for symmetric tensor field strength $G$ expressed in terms of a symmetric tensor potential, which we now write as $h_{\mu\nu}$, and view as a metric perturbation: $h= g-\eta$. Then $G$ has the interpretation as the linearized Einstein tensor:
\begin{equation}
G_{\mu\nu} = G^{(lin)}_{\mu\nu} \equiv R^{(lin)}_{\mu\nu} - \frac{1}{2}\eta_{\mu\nu} R^{(lin)}\, , 
\end{equation}
where $R^{(lin)}_{\mu\nu}$ is the linearized Ricci tensor and $R^{(lin)}$  its Minkowski trace.  
The equations (\ref{spin2}) can now be shown to be equivalent  to the linearized version of the one equation 
\begin{equation}\label{TMGlin}
G_{\mu\nu} +  \frac{1}{\mu} C_{\mu\nu}  =0 \, , 
\end{equation}
where $C_{\mu\nu}$ is the Cotton tensor (the 3D analog of the Weyl tensor): 
\begin{equation}\label{Cotton}
 \sqrt{-\det g}\ C_{\mu\nu} \equiv \varepsilon_\mu{}^{\tau\rho} D_\tau S_{\rho\nu}\, , \qquad 
 S_{\mu\nu} \equiv  R_{\mu\nu} - \frac{1}{4}g_{\mu\nu} R\, , 
\end{equation}
where $D$ is the usual covariant derivative constructed from the Levi-Civita connection. The tensor $S_{\mu\nu}$ is the 3D Schouten tensor, about which we shall have more to say later.  Equation (\ref{TMGlin}) can be derived from the Lagrangian density
\begin{equation}
{\cal L} =  -\sqrt{-\det g}\,  R + \frac{1}{\mu} {\cal L}_{LCS}\, , 
\end{equation}
where ${\cal L}_{LCS}$ is the Lorentz Chern-Simons (LCS) term.  This is the action of  ``topologically massive gravity''  (TMG) \cite{Deser:1981wh}.  Note the 
unconventional sign of the Einstein-Hilbert (EH) term; it  is needed for positive energy of the one massive spin 2 mode that is propagated in the Minkowski vacuum.

Applying the same procedure to the second-order  FP equations (\ref{FP}) for spin $2$, we arrive at the  Lagrangian density  for ``new massive gravity'' (NMG) \cite{Bergshoeff:2009hq}, which we may write in the form\footnote{The sign of the EH term depends on the metric signature convention; as we use here the ``mostly plus'' convention, the sign is opposite to that of \cite{Bergshoeff:2009hq} where the ``mostly minus'' convention was used. } 
\begin{equation}
{\cal L} = \sqrt{-\det g}\, \left[ -R + \frac{1}{m^2} K\right]  \, , \qquad K\equiv G^{\mu\nu}S_{\mu\nu}\, . 
\end{equation}
Note the occurrence, once again, of the Schouten tensor.  This  derivation of  NMG may be run in reverse to prove that it  propagates two massive spin-$2$ modes in a Minkowski vacuum, with helicities $\pm2$, but this fact does not guarantee that neither mode is a ghost (negative kinetic energy).  This is not an issue for TMG because one may always adjust the overall sign of the action to ensure that the one propagating mode is physical, but this may not be sufficient when there are two propagating modes.  In fact, if the same method is applied to the Proca equations then one arrives at an equivalent set of ``extended Proca''  equations \cite{Deser:1999pa}, but the ``extended Proca''  action  propagates one of the two spin-$1$ modes as a ghost, and the same phenomenon occurs for spin $3$ \cite{Bergshoeff:2009tb}. So the absence of ghosts in NMG is far from obvious. Nevertheless, there is an alternative form of the action involving an auxiliary tensor field that allows a simple proof of the equivalence of the linearized {\it action} to the standard  FP action, which is known to be ghost-free. This was reviewed in 
 \cite{Bergshoeff:2009fj}.  The absence of ghosts in linearized NMG may also be verified directly by a canonical analysis \cite{Deser:2009hb}.

The scalar $K= G^{\mu\nu} S_{\mu\nu}$ has the ``conformal covariance''  property \cite{Bergshoeff:2009hq}
\begin{equation}
g_{\mu\nu} \frac{\delta}{\delta g_{\mu\nu}} \int d^3 x \sqrt{|g|} K \quad \propto \quad K\, . 
\end{equation}
This implies that the quadratic approximation to $K$ is invariant under linearized Weyl transformations,  as stressed in \cite{Deser:2009hb}. In fact, the
scalar $G^{\mu\nu}S_{\mu\nu}$ has this property in any dimension.  To see this, one should first appreciate that the definition of the 
Schouten tensor is dimension dependent; for spacetime dimension $D>2$,  
\begin{equation}
S_{\mu\nu} =  \frac{1}{D-2} \left[ R_{\mu\nu} - \frac{1}{2(D-1)} R g_{\mu\nu} \right]\, . 
\end{equation}
This tensor first arose in the decomposition of the Riemann tensor into the traceless Weyl conformal tensor $W$  and a remainder:
\begin{equation}
R_{\mu\nu\rho\sigma} = W_{\mu\nu\rho\sigma} + \left(g\circ S\right)_{\mu\nu\rho\sigma}\, , 
\end{equation}
where $\circ$ indicates the Kulkarni-Nomizu product of two second-rank tensors (for symmetric tensors this product has the symmetries of the Riemann tensor). 
The Schouten tensor also has the interpretation as a (dependent) gauge potential for conformal boosts\footnote{This gauge potential is set equal to the Schouten tensor on imposing a constraint on conformal curvatures \cite{Kaku:1977pa} in close analogy to the way that the affine connection becomes a function of the metric and its derivatives when the torsion is constrained to vanish.} and this explains why, in 3D, the Cotton tensor may be expressed in terms of it.  
Next, we note the  {\it identity} \footnote{An equivalent identity has been noted independently in \cite{Oliva:2010zd}.}
\begin{equation}
R^{\mu\nu\rho\sigma} R_{\mu\nu\rho\sigma} -4R^{\mu\nu}R_{\mu\nu} + R^2 \equiv
W^{\mu\nu\rho\sigma}W_{\mu\nu\rho\sigma} -4(D-3) G^{\mu\nu} S_{\mu\nu}\, ,  
\end{equation}
which is valid for any $D\ge3$, although both sides vanish identically for $D=3$. For $D=4$ the left hand side is the integrand of the Gauss-Bonnet invariant, and hence is a total derivative. This shows  that the  4D scalar  $G^{\mu\nu}S_{\mu\nu}$ equals the square of the Weyl tensor, up to a total derivative. For $D>4$ the left hand side is the Lovelock term \cite{Lovelock:1971yv}; it is not a total derivative  but it has the feature that it does not contribute to the quadratic action in an expansion about Minkowski space\footnote{A number of other similarities between NMG  and  Lovelock gravity have been noted in 
 \cite{Sinha:2010ai, Oliva:2010eb, Myers:2010ru}.}.  Thus, even for $D>4$ it remains true  that  the quadratic approximation to $G^{\mu\nu}S_{\mu\nu}$ equals the square of the linearized Weyl tensor, up to a total derivative, and hence that it is invariant under linearized Weyl transformations. 

Let us now turn to the generalized, parity-violating, FP equations (\ref{genFP}). Applying the same procedure reviewed above for the other cases, we arrive at 
the Lagrangian density of  ``general massive gravity'' (GMG) \cite{Bergshoeff:2009hq}
\begin{equation}
{\cal L} = \sqrt{-\det g}\, \left[ -R + \frac{1}{m^2} G^{\mu\nu}S_{\mu\nu}\right]  + \frac{1}{\mu} {\cal L}_{LCS}
\end{equation}
where 
\begin{equation}
m^2 = m_+ m_-\, , \qquad \mu = \frac{m_+m_-}{m_- - m_+}\, .  
\end{equation}
Again, the derivation does not guarantee the absence of ghosts. However, the canonical analysis that shows NMG to be ghost free can be easily generalized to GMG  \cite{Andringa:2009yc}, as we now review. One begins by making a time-space split and imposing a convenient gauge condition 
\begin{equation}
\partial_i h_{i\mu} =0\, ;  \qquad \mu= (0,i) \quad i=1,2\, .
\end{equation}
The 3-metric perturbation $h$ can now be expressed in terms of three independent functions as follows:
\begin{equation}
h_{\mu\nu} =  \left(\begin{array}{cc} n & m\,  \hat\partial_i \phi_2\  \\ m\, \hat\partial_j \phi_2 \ & \hat\partial_i\hat\partial_j \phi_1\end{array}\right) \, ,  
\qquad \hat\partial_i \equiv \varepsilon^{ij} \partial_j  \, . 
\end{equation}
Substitution into the linearized GMG action yields an action involving $(n, \phi_1,\phi_2)$ that is fourth order in derivatives but only second-order in time derivatives. Introducing new independent functions $(N,\Phi_1,\Phi_2)$ by the {\it space} non-local  field redefinitions\footnote{Space non-local field redefinitions are allowed since they cannot  change the canonical structure, which depends on the time derivatives.}
\begin{equation}
n + \Box \phi_1 - 2m^2\left(\phi_1-\phi_2\right) = \frac{1}{\nabla^2} N 
\, , \qquad \phi_a = \frac{1}{\nabla^2} \Phi_a \quad (a=1,2), 
\end{equation}
we then arrive at an equivalent action with Lagrangian density
\begin{equation}\label{finres}
{\cal L} =  \frac{1}{2} \Phi_a \left[ \delta_{ab} \Box - M^2_{ab} \right] \Phi_b + 2m^2 N^2\, , 
\end{equation}
where the $2\times 2$ matrix $M^2_{ab}$ is real symmetric with eigenvalues $(m^2_+, m^2_-)$. This confirms that there are two propagating modes
with masses $m_\pm$. Crucially, both modes have positive energy, so GMG is ghost-free. 

It is remarkable that the final result (\ref{finres}) is  Lorentz invariant, despite the fact that we arrived at  it  via  manipulations that  explicitly violate Lorentz invariance, but the Lorentz transformations that leave invariant (\ref{finres}) are {\it not the same}  as the Lorentz transformations that leave invariant the 
linearized GMG action. This is simply a reflection of the fact that a free field theory has an infinite-dimensional invariance group. In the present case, this 
infinite dimensional group has at least two,  mutually space non-local and probably non-commuting, Lorentz subgroups. The introduction of interactions that preserve one of these Lorentz subgroups will  break the other one. Precisely because we view linearized NMG as the quadratic approximation to NMG, it is 
the manifest Lorentz transformations of this action, in its original form, that are relevant to the determination of the spin of  propagated modes.  For this reason, 
one cannot read off the spins  from the Lagrangian (\ref{finres}); however we  already know that the two modes have spin $2$ from the equivalence of the field equations to those of the generalized FP equations.

A feature of curvature-squared terms is that they contribute to the quadratic kinetic terms in an expansion about Minkowski space, and hence to the propagator 
as well as to the vertices of Feynmann diagrams. Specifically, they introduce $1/p^4$ type terms in the propagator, where $p$ is the 3-momentum, and this makes the generic curvature squared gravity theory power-counting renormalizable in 4D \cite{Stelle:1976gc}.  Unfortunately, this comes at the cost of unitarity. There is one exceptional case in which unitarity is not violated, although renormalizability is lost.  This is the model obtained by adding to the EH term  the square of the scalar curvature ($R^2$); this is equivalent, for an appropriate choice of signs, to a scalar field coupled to gravity (see \cite{Schmidt:2006jt} for a review).  There is another  exceptional case in which renormalizability is lost (without a gain in unitarity).  This is the model obtained by addition of the square of the Weyl curvature tensor ($W^2$), but without an $R^2$ term.  Exceptional cases can arise because  the  metric perturbation $h_{\mu\nu}$ is not an irreducible representation of the Lorentz group but includes the 
scalar trace $h$. There is a term in  the propagator that projects onto the pure spin 2 part of $h_{\mu\nu}$ and a term that projects onto the trace, irrespective of whether $h$ propagates any physical spin-$0$ mode.  Both terms in the propagator go like $1/p^2$ in the context of the EH term alone.  Addition of  the square of the Weyl conformal curvature  tensor  ($W^2$) causes  the  spin 2 projector part of the propagator to go like $1/p^4$ whereas addition  of  the square of the curvature scalar ($R^2$) causes the scalar projector part of the propagator to go like $1/p^4$, and {\it both} are needed for power-counting renormalizability.  
Thus, omitting either the $R^2$ term or the $W^2$ term implies a loss of renormalizability. 

The situation in 3D, where the generic curvature-squared gravity theory is power-counting super-renormalizable, is potentially better since at least the  $K=G^{\mu\nu}S_{\mu\nu}$ term may be included without violating unitarity, as we have just seen. However, the linearized Weyl  invariance of  $K$ implies  that the $1/p^2$ behaviour of the scalar projection term in the propagator is not affected by the curvature-squared term of NMG,  so that NMG is not power-counting renormalizable  \cite{Deser:2009hb}. Here it should be said that  it has been claimed that NMG is  super-renormalizable \cite{Oda:2009ys},  but  we have not understood the argument\footnote{In particular, it appears that the means used to arrive at this conclusion are not specific to 3D and could  be used to obtain the same result in 4D.}. In any case, it is certainly true that NMG is exceptional within the class of curvature-squared theories in 3D in much the same way that 
the `$R+W^2$' theory is exceptional in 4D. This can be seen more explicitly using the results of  \cite{Nishino:2006cb} for the propagator of the general 3D gravity model with curvature squared terms. Consider the Lagrangian density
\begin{equation}
{\cal L} = \sqrt{|g|} \left[ \sigma R + \frac{a}{m^2} K + \frac{b}{m^2} R^2 \right] 
\end{equation}
for constants $(\sigma,a,b)$; the choice $(\sigma,a,b) =(-1,1,0)$ yields NMG. If we expand about the Minkowski vacuum we find that 
\begin{equation}
{\cal L} = \frac{1}{2} h^{\mu\nu} {\cal O}_{\mu\nu ,\rho\sigma} \, h^{\rho\sigma} + \dots
\end{equation} 
where  ${\cal O}$ is a fourth-order linear differential tensor operator and the dots indicate  interaction terms. The operator ${\cal O}$ may be expressed 
in terms of two orthogonal projection operators, for spin 2 and  spin 0  \cite{VanNieuwenhuizen:1973fi}; in momentum space, these are
\begin{equation}
P^{(2)}_{\mu\nu , \rho\sigma} = \frac{1}{2}\left( \theta_{\mu\rho} \theta_{\nu\sigma} +  \theta_{\mu\sigma} \theta_{\nu\rho} 
- \theta_{\mu\nu}\theta_{\rho\sigma}\right)\, , \qquad P^{(0,s)}_{\mu\nu ,\rho\sigma} = \frac{1}{2} \theta_{\mu\nu} \theta_{\rho\sigma}\, , 
\end{equation}
where
\begin{equation}
\theta_{\mu\nu} = \eta_{\mu\nu} - \frac{p_\mu p_\nu}{p^2}\, . 
\end{equation}
Specifically, one finds that
\begin{equation}
{\cal O}_{\mu\nu , \rho\sigma} =\left[-\frac{1}{2} \sigma p^2 + \frac{ap^4}{2m^2} \right] P^{(2)}_{\mu\nu ,\rho\sigma} +  \left[ \frac{1}{2} \sigma p^2 + \frac{bp^4}{2m^2}\right] P^{(0,s)}_{\mu\nu ,\rho\sigma}\, . 
\end{equation}
This operator is not invertible, but we may invert within each of the subspaces defined by the two projectors. The result is the propagator 
\begin{equation}
2m^2 \left\{ \frac{P^{(2)}_{\mu\nu , \rho\sigma}}{ p^2\left(ap^2 -m^2\sigma\right)}  + 
\frac{P^{(0,s)}_{\mu\nu , \rho\sigma}}{p^2\left(bp^2 + m^2\sigma\right)}\right\}\, .  
\end{equation}
As long as $ab\ne0$ this behaves like $1/p^4$ as $p^2\to\infty$, but the spin 2 part goes like $1/p^2$ when $a=0$ and the spin 0 part goes like $1/p^2$ when $b=0$.

Returning to 4D, there is one curvature-squared model that is potentially renormalizable, and that is conformal gravity. The action for conformal gravity is just the integral of $W^2$, without the EH term. The propagator is now purely $1/p^4$ because the trace of the metric perturbation is a gauge degree of freedom. 
It is often said that this model has ghosts since any perturbation away from conformality leads to  ghosts, but  consistency  requires that the conformal invariance be preserved, even  by quantum corrections.  The one-loop conformal anomalies cancel for some conformal supergravity models (see \cite{Fradkin:1985am} for a review) so these may be viable theories, although they have not yet found any compelling application.  

The action for 3D conformal gravity is  just the LCS term \cite{vanNieuwenhuizen:1985cx}. In other words, one omits the EH term from TMG, but this  propagates no modes.  One may also omit the  EH term from NMG, in which case one gets a model that propagates a single massless mode \cite{Deser:2009hb}, of no definite spin because spin is not defined for massless particles. If a LCS term is added (equivalently, if the EH term is omitted from the GMG action) then  one gets a 4th order ``New Topologically Massive Gravity'' (NTMG) model that propagates a single massive spin-$2$ mode \cite{Andringa:2009yc,Dalmazi:2009pm}. 
In any of these models without an EH term, the trace of the metric perturbation is a gauge degree of freedom in the quadratic approximation, so the  propagator is either pure $1/p^4$ or behaves this way  in the short distance limit. This fact was claimed in \cite{Deser:2009hb} to imply renormalizabilty. However, the trace of  the metric perturbation is  not a gauge degree  of freedom  of the interacting theory. Its equation of motion is $K=0$, which is identically satisfied in the linearized limit (since $K$ has no term linear in fields) but  not  otherwise. The usual power-counting arguments apply to a perturbation theory in which all non-gauge degrees of freedom are represented in the propagator, but this condition is not satisfied here.  It remains to be seen what effect this has. 

If massive gravity theories are not power-counting renormalizable then they are still no worse than general relativity in 4D. In the latter case, we know that supersymmetry can soften the ultra-violet divergences, and there are some hints that the maximally supersymmetric ${\cal N}=8$ supergravity may be finite (see e.g. \cite{Dixon:2010gz}). In view of this, it is of  interest to consider the massive 3D supergravity theories. The   representation theory for massive particle supermultiplets in 3D is formally the same as that for massless supermultiplets in 4D so  ${\cal N}=8$ is maximal for massive 3D supergravity too, although  the total number of supersymmetry charges is 16 rather than 32.  So far only the ${\cal N}=1$ theory has been constructed in detail \cite{Andringa:2009yc,Bergshoeff:2010mf} although the ${\cal N}=2$ massive 3D supergravity has been constructed as a linear theory in Minkowski space  \cite{Andringa:2009yc} and it appears as though the ${\cal N}=8$ theory can be constructed in the same approximation \cite{Bergshoeff:2010ui}.  Assuming that there is an  ${\cal N}=8$  massive 3D supergravity, representation theory implies that it must preserve parity since the state of helicity $+2$ is in the same supermultiplet as the state of helicity $-2$.  In other words, we expect NMG to have an ${\cal N}=8$  supersymmetric extension but not TMG. The representation theory would allow ${\cal N}=7$ as maximal for super-TMG but the details suggest that ${\cal N}=6$ is actually maximal for a parity violating massive 3D gravity. 
Thus, ${\cal N}=8$ new massive supergravity is the most promising candidate for a 3D massive gravity theory with `improved'  ultraviolet behaviour, but 
it remains to be seen how significant any improvement will be.

%%%%%%%%%%%%%%%%%%%%

\subsection*{Acknowledgments}
PKT thanks the conference organizers for the invitation to deliver this talk on the occasion of  Emilio Elizalde's 60th birthday. The authors
are grateful to Roel Andringa, Mees de Roo, Jan Rosseel and Ergin Sezgin for discussions in the course of collaborations on supersymmetric extensions 
of massive 3D gravity models.  The work of OH is supported by the DFG -- The German Science Foundation
and in part by funds provided by the U.S. Department of Energy (DoE) under the cooperative research agreement DE-FG02-05ER41360. PKT thanks the 
EPSRC for financial support.

\end{document}